\documentstyle[12pt,epsf]{article}
 \font\blackboard=msbm10 
 \font\blackboards=msbm7 \font\blackboardss=msbm5
 \newfam\black \textfont\black=\blackboard
 \scriptfont\black=\blackboards \scriptscriptfont\black=\blackboardss
 \def\Bbb#1{{\fam\black\relax#1}}

\font\ninerm=cmr9
\def\uniset{\rlap{\ninerm 1}\kern.15em 1}
\def\e{\mathop{\rm e}\nolimits}

\title{Eigenstate structures around a hyperbolic point}
\author{{\bf S. Nonnenmacher} and {\bf A. Voros}\\
\\
CEA--Saclay, Service de Physique Th\'eorique\\
F-91191 Gif-sur-Yvette CEDEX (France)}
\begin{document}

\maketitle
\abstract
Using coherent-state representations of quantum mechanics
(Bargmann, Husimi, and stellar representations), 
we describe analytically the phase-space structure of the general eigenstates
corresponding to a 1-dimensional bilinear hyperbolic Hamiltonian, $H=pq$ or
equivalently $H={1\over 2}(P^2-Q^2)$. Their semi-classical behaviour is discussed
for eigenvalues either near or away from the separatrix energy $\{H=0\}$,
especially in the phase-space vicinity of the saddle-point $(q,p)=(0,0)$.

\section{Introduction}
Semiclassical theory, and the quantum--classical correspondence, 
are still incompletely understood at
the level of long-time or invariant structures, especially when the classical
dynamics shows exponential sensitivity to initial conditions (instability,
or positive Lyapunov exponents).

We have therefore selected a simple 1-d system, the dilation operator
which quantizes the classical linear hyperbolic Hamiltonian $H(q,p)=qp$,
and collected some detailed analytical properties of its eigenstates.
Elementary as this quantum Hamiltonian may be,
seemingly unreported expressions for its eigenstates are given here,
precisely within the coherent-state (Bargmann or Husimi) formulations
where the semiclassical behaviour ($\hbar \to 0$ asymptotics) of quantum states 
is best seen.

Formulae for this ``dilator" should be useful tools to probe 
quantum phenomena linked to unstable (here, hyperbolic) classical dynamics.
Here are two actively studied examples involving this Hamiltonian: 

- in one dimension, classical hyperbolic dynamics takes place 
near a saddle-point (assuming it is generic, i.e., isolated and nondegenerate);
then it is locally equivalent to $H=pq$ in suitable canonical coordinates,
i.e, the classical dilator is the local normal form for this class of problems.
At the quantum level, 1-d saddle-points also challenge semiclassical analysis:
like energy minima, they correspond to critical energy values 
(i.e., the phase-space velocity vanishes);
semiclassical analysis is fundamentally harder at critical
than at regular energy values (where simple WKB theory works), 
but saddle-points are even harder to understand than minima
and their study has expanded more recently \cite{CdV,bleher,paul}. 

- in higher dimensions, 
the search for quantum manifestations of unstable classical motion
forms one facet of the ``quantum chaos" problem.
For instance, the role and imprint of unstable periodic orbits upon quantum
dynamics remain actively debated issues. 
It is known that they influence both the quantum bound state
energy spectrum (through trace formulas) and wave functions (through scarring).
More specifically, the (stable) orbits of elliptic type generate
quantization formulas and quasimode constructions in a consistent way.
In chaotic systems, by contrast, 
trace formulas diverge and scarring occurs rather unpredictably, so that
periodic orbits (now unstable) have incompletely assessed quantum effects;
nevertheless, an essential role must still go to their linearized classical
dynamics, which is of hyperbolic type hence generated by linear dilator(s).

Fully chaotic behaviour, a combination of global instability plus
ergodic recurrence, cannot however be captured by integrable models;
therefore, the place of the 1-d dilator eigenfunctions in quantum structures
corresponding to fully chaotic dynamics remains to be further studied.
At present, these eigenfunctions should primarily find use as microlocal models
for general 1-d eigenstates near saddle-points
(and for separatrix eigenstates in higher-dimensional integrable systems,
by extension).

\section{Coherent-state representations}
\subsection{Bargmann representation}
The Bargmann representation \cite{barg} is a particular coherent-state representation
of quantum wave-functions \cite{klau,perel,gilmore} in terms of entire functions. 
Although it can be defined in any dimension, 
we will just use it for 1-dimensional problems;
it then transforms Schr\"{o}dinger wave-functions $\psi(q)$
defined over the whole real line 
into entire functions $\psi(z)$ of a complex variable $z$, as 
\begin{equation}
\psi(z)\;=\;\langle z|\psi\rangle \;=\;{1\over(\pi\hbar)^{1/4}} \int_\Bbb R
\e^{{1\over \hbar}(-{1\over 2}(z^2+q'^2)+\sqrt{2}zq')}\;\psi(q')\;dq'.
\end{equation}
Here $|z\rangle $ denotes a (Weyl) coherent state localized at the phase space point
$(q,p)$ where $z=2^{-1/2}(q-ip)$, and satisfying $\langle z|z'\rangle =\e^{z\overline {z'}/\hbar}$;
i.e. these coherent states are not mutually orthogonal, 
and their normalization is not $\langle z|z\rangle =1$, allowing instead
the bra vector $\langle z|$ to be a holomorphic function of its label $z$.
On the other hand, a closure formula exists 
which makes the Bargmann transform invertible,
\begin{equation}
\label{clos}
\uniset = \int_\Bbb C {d\Re(z)d\Im(z)\over\pi\hbar}\;
\e^{-{z\overline z/\hbar}}\,|z\rangle \langle z|.
\end{equation}

The Bargmann transformation maps ordinary square-integrable wave-functions
into a Hilbert space of entire functions of order $\le 2$.
We will however mostly deal with generalized wave-functions $\psi(q)$,
which are not $L^2$ but only tempered distributions.
They can then still be Bargmann-transformed by the integral formula (1),
and into entire functions of order $\le 2$ as before, now bounded as
\begin{equation}
|\psi(z)| \leq c(1+|z|^2)^N \e^{|z|^2\over 2\hbar} \qquad \mbox{for some } N.
\end{equation}
This in turn constrains the distribution of their zeros \cite{boas}: 
the counting function
\begin{equation}
n(r)=\#\{\mbox{zeros }\,z_m\,\mbox{ of }\,\psi(z)\,\mbox{ s.t. }\,|z_m|\leq r\}
\end{equation}
(zeros will be always counted with their multiplicities) verifies
\begin{equation}
\limsup_{r\to\infty}\, {n(r)\over r^2} \leq {\e\over\hbar}.
\end{equation}
It follows that a Bargmann function admits a canonical Hadamard representation
as an (in)finite product over its zeros,
\begin{equation}
\label{prod}
\psi(z) = \e^{p(z)} \,z^{n(0)} \,  \prod_{z_m \ne 0} \Bigl(1-{z\over z_m}\Bigr)
\exp \Bigl({z\over z_m} + {1 \over 2}\Bigl({z\over z_m}\Bigr)^2\Bigr)
\end{equation}
where $p(z)=\hbar^{-1}(a_2z^2 + a_1z + a_0)$, with moreover $|a_2|\leq 1/2$; 
the integer $n(0)\ge 0$ is the multiplicity of $z=0$ as a zero of $\psi(z)$.
This decomposition shows that the wave-function is completely determined by 
the knowledge of all the Bargmann zeros (with their multiplicities) plus three
coefficients $a_2,a_1$, and $a_0$ (which only fixes a constant factor).

When the Hamiltonian is a polynomial (or more generally an analytic function) 
in the variables $(q,p)$, it can be useful to express its quantum version 
as a pseudo-differential operator acting on the Bargmann function \cite{barg},
using the rules
\begin{eqnarray}
a^\dagger = {\hat q-i\hat p\over\sqrt2}\ \ \qquad\mbox{(creation operator)}
&\rightarrow& \mbox{multiplication by } z\nonumber\\
  a  = {\hat q+i\hat p\over\sqrt2} \quad\mbox{(annihilation operator)}
&\rightarrow& \hbar{\partial\over\partial z}.
\end{eqnarray}

\subsection{Husimi representation}
An alternative point of interest lies in certain semi-classical densities 
on phase space associated to wave-functions.
In particular,  the Wigner function is defined as
\begin{equation}
{\cal W}_\psi (q,p) = (2 \pi \hbar)^{-1} \int_\Bbb R 
\psi(q-r/2) \overline{\psi}(q+r/2) \e^{ipr/\hbar} dr
\end{equation}
and the Husimi function \cite{husimi} as the convolution of the Wigner function 
with a phase-space Gaussian,
\begin{equation}
{\cal H}_\psi (q,p) =(\pi \hbar )^{-1} \int_{\Bbb R ^2} {\cal W}_\psi(q',p')
 \e^{- \left[(q-q')^2+
(p-p')^2 \right]/ \hbar} dq'dp'.
\end{equation}
The Wigner representation has greater symmetry (invariance under all linear 
symplectic transformations), 
but Wigner functions show a much less local semiclassical behaviour: 
they tend to display huge nonphysical oscillations, 
which must be averaged out to reveal any interesting limiting effects.
In the Husimi functions, the spurious oscillations get precisely damped so as
to unravel the actual phase-space concentration of the semiclassical measures,
but at the expense of reducing the invariance group.
The Husimi function is equivalently given by
\begin{equation}
\label{hus}
{\cal H}_\psi(z,\overline z)\,=\,{\langle z|\psi\rangle \langle \psi|z\rangle 
\over \langle z|z\rangle }\,=\,|\psi(z)|^2\,\e^{-z\overline z/\hbar}
\end{equation}
hence it constitutes the density of a positive measure on the phase space;
for the scattering-like eigenfunctions to be studied here, 
this measure will not be normalizable.
(For a normalized state, it is a probability measure thanks to the closure
formula (\ref{clos}).)

It is interesting to study how the Husimi measure of an eigenfunction 
behaves as $\hbar\to 0$. For an energy away from the separatrix, 
a standard theorem states that this measure concentrates on the
energy surface, with a Gaussian transversal profile \cite{taka,kurchan,vor89}. 
For energies close to a separatrix, careful analyses were performed in 
\cite{CdV, bleher,paul}. 
Our aim here is to select a simple tractable case, 
namely the eigenfunctions of the linear hyperbolic Hamiltonian \cite{taka:hyp},  and to carry further its description by means of the Bargmann representation, 
using eq. (\ref{hus}) to derive the Husimi density as a by-product.

\subsection{Stellar representation}

According to the factorized representation (\ref{prod}),
1-d quantum wavefunctions can be essentially parametrized 
in a phase-space geometry by the distribution of their Bargmann zeros which,
by eq. (\ref{hus}), is also the pattern of zeros for the Husimi density itself;
it thus constitutes a complementary viewpoint to the previous emphasis put on
the high-density behaviour of the Husimi function.
We refer to this ``reduced" description of a wavefunction by a discrete cloud
of phase-space points as a stellar representation.
It puts quantum mechanics in a new perspective \cite{leb:vor,leb,tualle},
but calls for a finer understanding of both dynamical and asymptotic properties 
of Bargmann zeros if new results are to be awaited through eq. (\ref{prod}).
Consequently, our subsequent analysis of ``toy"
 eigenfunctions
will largely deal with explicit behaviours of their Bargmann zeros.

\section{Description of the framework}

\subsection{The linear hyperbolic Hamiltonian}
The classical 1-dimensional Hamiltonian of linear dilation is $H(q,p)=pq$,
which is also equivalent to the scattering Hamiltonian $H={1\over2}(P^2 - Q^2)$ 
upon a symplectic rotation of the coordinates by $\pi/ 4$ according to
\begin{equation}
q={P+Q \over\sqrt{2}}, \qquad p={P-Q \over\sqrt{2}}.
\end{equation}
A classical trajectory at any energy $E \ne 0$ is a hyperbola branch,
\begin{equation}
q(t) =q_0\, \e^t, \qquad  p(t) =p_0\,\e^{-t} \qquad \mbox{with } E=p_0 q_0
\end{equation}
whereas the $E=0$ set is a separatrix, made of a stable manifold $\{p=0\}$,
an unstable manifold $\{q=0\}$, and the hyperbolic fixed point $(0,0)$.

We study eigenfunctions of the operator obtained by Weyl quantization,
namely $\hat H = {\hbar\over i}(q{d\over dq}+{1\over 2})$.
This quantum Hamiltonian admits two independent stationary wave-functions 
for any real energy $E$:
\begin{equation}
\psi_\pm^E = K\,\theta(\pm q)\,\e^{(i{E\over \hbar}-{1\over 2})\log |q|}
\end{equation}
where $\theta(q)$ is the Heaviside step function; $K\ne 0$ is a complex constant
(having no preferred value, since the solutions are not square-integrable). 
Microlocally, each of these wavefunctions is supported by the lagrangian
manifolds $\Lambda_\pm^E =\{pq=E\,, \pm q>0\,\}$, i.e. half of the $E$-energy
surface \cite{CdV}. 

In order to obtain more semi-classical information, we will use the Bargmann representation. For instance,
\begin{equation}
\label{int}
\langle z|\psi_+^E\rangle =\psi_+^E(z) = {K\over (\pi\hbar)^{1/4}}\int_0^\infty
\e^{{1 \over \hbar}(-{1 \over 2}(z^2+q'^2)+\sqrt{2}zq'+iE\log q' )}
{1\over \sqrt {q'}}dq'
\end{equation}
and we have of course 
\begin{equation}
\label{sym}
 \psi_-^E(z) = \psi_+^E(-z); \qquad  
\psi_\pm^{-E}(z) = \overline {\psi_\pm^E(\overline z)}.
\end{equation}
Our aim in this paper is to describe the general eigenfunction of energy $E$
in this representation: up to a global (removable) constant factor, 
it reads as $\psi_\lambda^E(z)=\psi_+^E(z)+\lambda \psi_-^E(z)$, 
for any complex projective parameter $\lambda$, i.e. 
$\lambda \in \overline \Bbb C = \Bbb C \cup\{\infty\}$.
We can immediately restrict attention to $E \ge 0$ due to the second of
eqs. (\ref{sym}).
We will be particularly interested, 
on the one hand, in the global profile of these functions, 
and on the other hand, in the position of their zeros 
because these form the main skeleton of
the Hadamard product representation (\ref{prod}) \cite{leb:vor}. 

The motivation is to better describe the eigenfunctions of a general 
1-d Hamiltonian for eigenvalues close to a classical saddle-point energy value.
Such an eigenfunction cannot be simply of WKB form near the saddle-point;
instead, it should be microlocally modeled by an eigenvector $\psi_\lambda^E$
of the dilation operator $\hat H$ near $(q,p)=0$ with $E \approx 0$
(up to straightforward displacements, in phase space and in energy).
Thus, in a Bargmann representation, the eigenfunction near $z=0$ 
ought to behave like $\psi_\lambda^E(z)$ for some $\lambda$, which is the one
quantity whose actual value is determined by global features of the solution.
(In particular, for a parity-symmetric 1-d Hamiltonian
and for a saddle-point located at the symmetry center, 
only even or odd solutions ever come into play;
hence parity conservation preselects the two eigenfunctions of $\hat H$ 
with the special values $\lambda=+1$ and $-1$ respectively.)

\subsection{Main analytical results}
The functions $\psi_\lambda^E(z)$ are also solutions
of the Schr\"odinger equation written in the Bargmann representation,
\begin{equation}
\label{bar}
{i\over2}\Bigl(-\,\hbar ^2 {d^2\over dz^2}\,+\,z^2\Bigr)\psi^E(z) =E\:\psi^E(z).
\end{equation}
It is convenient to use the rotated variables 
$Q,P$ and $Z = 2^{-1/2}(Q-iP) =z \e^{-i\pi /4}$ in parallel with $q,p$, and $z$.
In those variables the quantum Hamiltonian reads as 
the quadratic-barrier Schr\"odinger operator 
$\hat H = {1\over2}(-\hbar^2 d^2/dQ^2 - Q^2)$.
Its Bargmann transform happens to be exactly the same operator in the $Z$
variable, simply continued over the whole complex plane, so that
the eigenfunction equation can also be written as
\begin{equation}
\label{barrier}
{1\over2}\Bigl(-\,\hbar^2 {d^2\over dZ^2}\,-\,Z^2\Bigr)\Psi^E(Z) = E\:\Psi^E(Z).
\end{equation}
At the same time, the Bargmann representations obtained from $q$ and $Q$ 
are equivalent under a simple complex rotation,
\begin{equation}
\Psi(Z) \,=\,\psi(z) \qquad \mbox{with } Z=z \e^{-i\pi /4} .
\end{equation}

Consequently, as a main first result, the above solutions are directly related
to the parabolic cylinder functions $D_{\nu}(y)$, defined for example in \cite{bat}.
As a matter of fact, we have :
\begin{equation}
\label{psi-D}
\Psi_\pm ^E(Z)= \psi_{\pm}^E(z) 
= K{\hbar^{iE\over 2\hbar} \over {\pi ^{1/4}}}\, 
\Gamma \left({1\over 2}  + {iE \over \hbar}\right)\, 
D_{- {1\over 2} -{iE \over \hbar}}\biggl(\mp\sqrt{2\over \hbar} z\biggr).
\end{equation}
Up to rescaling, the Bargmann eigenfunction
$\psi_\lambda^E(z)$ is then simply a linear combination of known functions,
$[\lambda D_{\nu}(y)+D_{\nu}(-y)]$ with the notations 
$y=\sqrt{2/\hbar}\,z$ and $\nu = -{1\over 2}-i{E\over \hbar}$. 

The situation simplifies even further on the separatrix $E=0$,
where parabolic cylinder functions reduce to Bessel functions, as
\begin{eqnarray}
\label{bessel}
D_{-1/2}(y) &=& \left((2\pi)^{-1} y \right)^{1/2}
K_{1/4}\left({y^2 /4}\right) \nonumber\\
D_{-1/2}(y)\,+\,D_{-1/2}(-y) &=& \e^{+i\pi /8}\, (\pi  y)^{1/2}
J_{-1/4}\left(i{y^2 /4}\right)\\
-D_{-1/2}(y)\,+\,D_{-1/2}(-y) &=& \e^{-i\pi /8}\, (\pi y)^{1/2}
J_{1/4}\left(i{y^2 /4}\right). \nonumber
\end{eqnarray}

By virtue of eq. (\ref{hus}), eqs. (\ref{psi-D}) and (\ref{bessel}) 
yield the Husimi densities in closed form, for all eigenfunctions;
e.g., for $\psi_+^E (z)$ and $\Psi_{\pm 1}^0 (Z)$ respectively,
\begin{eqnarray}
\label{husi}
{\cal H}_+^E (z,\overline z) &=& {|K|^2 \sqrt \pi \over \cosh (\pi E/\hbar)}
|D_{-{1\over 2}-i{E\over\hbar}}(-\sqrt{2/\hbar}\,z)|^2 \e^{-z\overline z/\hbar} 
 \\
{\cal H}_{\pm 1}^0 (Z,\overline Z) &=& |K|^2 \sqrt{2 \pi^3/\hbar} 
\,|Z|\,|J_{\mp 1/4} (Z^2 /\, 2 \hbar)|^2 \e^{-Z\overline Z/\hbar} . \nonumber 
\end{eqnarray}
This is an interesting extension of earlier similar formulas for
the Wigner functions; e.g., for $\psi_+^E$, \cite{bal:vor}
\begin{eqnarray}
{\cal W}_+^E (q,p) &=& {|K|^2 \over \hbar \,\cosh (\pi E/\hbar)}\theta(q)
\e^{2ipq/\hbar} {_1 F_1}\Bigl({1\over 2}+{iE \over \hbar};1;-4ipq/\hbar\Bigr) \\
{\cal W}_+^{E=0}(q,p) &=& |K|^2 \hbar^{-1} \theta(q)\,J_0(4ipq/\hbar) 
\nonumber
\end{eqnarray}
where ${_1 F_1} (-\nu;1;x) \propto L_\nu (x)$ (Laguerre functions).
These expressions are specially simple (constant along connected orbits)
because the Wigner representation is exactly $\hat H$-invariant 
under the symplectic evolution generated by the bilinear Hamiltonian $H$.
Inversely, coherent-state representations can never preserve 
such a dynamical invariance of hyperbolic type,
hence closed-form results like (\ref{husi}) are necessarily more intricate
than their Wigner counterparts and not simply transferable therefrom.

Fig. 1 shows contour plots for some of those Husimi densities, 
all with the normalizations $K=\hbar=1$.
The first example (${\cal H}_+^E(z,\overline z)$ for $E=+1$) is plotted twice: 
once with equally spaced contour levels starting from zero (linear scale),
and once with contour levels in a geometric progression decreasing
from the maximum (logarithmic scale). 
The linear plot emphasizes the high-density modulations which control
the measure concentration of the Husimi density;
the logarithmic plot reveals the subdominant structures and especially
the locations of the zeros $z_m$ 
which determine the Hadamard parametrization. 
For the parabolic cylinder function itself as an example,
the factorization formula (\ref{prod}) specifically gives
\begin{eqnarray}
D_\nu(z) &=& \e^{p(z)}  \prod_{z_m \ne 0} \Bigl(1-{z\over z_m}\Bigr)
\exp \Bigl({z\over z_m}+{1 \over 2}\Bigl({z\over z_m}\Bigr)^2\Bigr), \nonumber\\
p(y) &\equiv& 
\log D_\nu(0)+(\log D_\nu)'(0)\ y+(\log D_\nu)''(0)\ y^2/2 \nonumber\\
&=& 
-\log { \Gamma \bigl({1-\nu \over 2}\bigr) \over 2^{\nu/2} \sqrt \pi} 
-\sqrt 2\ 
{\Gamma\bigl({1-\nu\over 2}\bigr)\over\Gamma\bigl(-{\nu\over 2}\bigr)}\ y
-\left( {\nu \over 2} + {1\over 4}+
\left[{\Gamma\bigl({1-\nu\over 2}\bigr)\over\Gamma\bigl(-{\nu\over 2}\bigr)}
\right]^2
\right) y^2 .
\end{eqnarray}
In order to save figure space, we do not provide the log-plots 
used to locate the zeros for the other Husimi densities 
but only their linear contour plots, with the zeros superimposed as small dots.
(The same uniform contour level spacing is used throughout 
to make comparisons easier.)

\section{Asymptotic expansions}
We can then rely upon the known asymptotic properties of the $D_{\nu}(y)$,
which follow from the integral representation (\ref{int}),
to investigate two asymptotic regimes for the eigenfunctions of $\hat H$. 
Firstly, when $E/\hbar$ is kept finite, the
asymptotic expansions will be valid in the limit $y\to\infty$; this corresponds
to eigenenergies very close to the classical separatrix energy $E=0$. Secondly, 
if we fix the energy $E$ at a non-vanishing value and let $\hbar \to 0$, we
have to use a different type of asymptotics, namely usual WKB expansions. 

\subsection{Energies close to zero}
 We use the expansions for $D_\nu (y)$ when $|y|\to\infty$ \cite{bat} at fixed $\nu$,
which are obtained from integral representations using Watson's lemma, 
and take different forms in various angular sectors, 
\begin{eqnarray}
\label{dexp}
-{\pi \over 2}<\arg(y)<+{\pi \over 2}&:&\quad D_\nu(y) \sim y^\nu \e^{-y^2/4}
\sum_{n\ge 0} {\Gamma(2n-\nu)\over n!\ \Gamma(-\nu)}{1\over(-2y^2)^n}
\sim y^\nu \e^{-y^2/4}(1+O(y^{-2})) \\
-{\pi}<\arg(y)<-{\pi \over 2}&:&\quad D_\nu(y)\sim y^\nu\, \e^{-y^2/4}
(1+O(y^{-2}))\,\,-{\sqrt{2\pi}\over \Gamma(-\nu)}\e^{-i\nu\pi}\,y^{-\nu-1}\,
\e^{y^2/4}\,(1+O(y^{-2}))\nonumber\\
+{\pi \over 2}<\arg(y)<{+\pi }&:&\quad D_\nu(y)\sim y^\nu\, \e^{-y^2/4}
(1+O(y^{-2}))\,\,-{\sqrt{2\pi}\over \Gamma(-\nu)}\e^{i\nu\pi}\,y^{-\nu-1}\,
\e^{y^2/4}\,(1+O(y^{-2})). \nonumber
\end{eqnarray}
The sectors are specified here as non-overlapping and bounded by Stokes lines,
i.e. curves of maximal dominance of one exponential factor over the other.
(Each asymptotic expansion actually persists in a larger sector
overlapping with its neighbours, but this extension will not be of use here.) 
The above expansions are valid for $|y|\to\infty$ within each sector
and provide approximations to the shape of the eigenfunctions 
and the positions of their zeros for large $|y|$; 
we will use them  to leading order only (up to $O(y^{-2})$ terms). 
For the general eigenfunction 
$\psi^E_\lambda \propto [\lambda D_{\nu}(y)+D_{\nu}(-y)]$,
eqs. (\ref{dexp}) straightforwardly generate four different expansions 
in the four $z$-plane quadrants $S_j,\ j=0,1,2,3$ 
(named anticlockwise from $S_0=\{0<\arg z <\pi/2\}$).

\subsection{WKB expansions for a fixed non-vanishing energy}
The previous expansions are inapplicable when $\hbar\to 0$ 
with the classical energy kept fixed at a non-zero value (in the following we
will suppose $E>0$). In this regime, we have to use WKB-type expansions instead.
These can be obtained from the integral representations of the solutions
(\ref{int}), by performing saddle-point approximations ; 
equivalently, they can be found directly from
the Schr\"odinger equation written in Bargmann variables (\ref{barrier}). 
We will use the $Z$ variable for convenience. 

The general WKB solution can be written, to first order in $\hbar$, as
\begin{equation}
\Psi ^E(Z) \sim (2E+Z^2)^{-1/4} \left(\alpha(\hbar) \e^{+i\phi(Z_0,Z)/\hbar}
+ \beta(\hbar) \e^{-i \phi(Z_0,Z)/\hbar} \right), \qquad
\end{equation}
where the exponents are now the classical action integrals, 
taken from an (adjustable) origin $Z_0$,
\begin{equation}
\phi(Z_0,Z) = \int_{Z_0}^Z  P_E(Z')dZ', \qquad P_E(Z) \equiv \sqrt{2E+Z^2}
\end{equation}
with the determination of the square root $P_E(Z)$ fixed 
by the cuts indicated on fig. 2 (left) and by $P_E(0)>0$.
This approximation is valid for $\hbar\to 0$, 
when $Z$ stays far enough from the two turning points 
$Z_\pm=\pm i\sqrt{2E}$ in the sense that $|\phi(Z_\pm,Z)| \gg \hbar$.
 The coefficients $\alpha$ and $\beta$ a priori depend
upon $\hbar$ and the region of the complex plane where $Z$ lies. 
More precisely, the complex $Z$-plane is to be partitioned by the Stokes lines,
specified for each turning point by the condition
\begin{equation}
i \phi(Z_\pm,Z)/\hbar\quad \mbox{real} .
\end{equation}
Three such lines emanate from every turning point.
When the variable $Z$ crosses a Stokes line, the coefficients $\alpha$ and $\beta$ change according to connection rules (see \cite{olver} for instance); 
the application of these rules yields the global structure of the solution. 
For full consistency, no Stokes line should link two turning points; 
this restriction forces us to slightly rotate $\hbar$ into the complex plane, 
as $\hbar\to \e^{i\epsilon}\hbar$, 
with the resulting partition of the plane drawn on fig. 2 (left).
The explicit form of exponentially small WKB contributions 
is generally sensitive to the choice $\epsilon = \pm 0$. 
However, such is not the case for the subsequent results at the order to which they will be expressed, so that we may ultimately reset $\epsilon =0$.

Before studying a particular solution, we introduce the hyperbolic angle variable $\theta = {\rm arcsinh} (Z/\sqrt{2E})$ which allows 
to integrate the action in closed form, as
\begin{equation}
\phi(Z_0,Z)={E\over 2} \left[\sinh(2\theta')+2\theta'\right]_{\theta_0}^{\theta}
={1 \over 2} \left[ Z'\sqrt{2E+Z'^2} + \log (Z'+ \sqrt{2E+Z'^2})\right]_{Z_0}^Z;
\end{equation}
the turning points correspond to $\theta_\pm = \pm i\pi/2$;
the action values $\phi(0,Z_\pm)=\pm i \pi E/2$ are frequently needed.

We fully describe one eigenfunction as an example, $\Psi_+^E(Z)$ 
(corresponding to $\lambda=0$).
We identify its WKB form first in the regions ${\cal S}'$, ${\cal S}_1$, ${\cal S}_2$,
by noticing that this eigenfunction must be exponentially decreasing 
for $Z \to \infty$ in a sector around $\arg Z = -3\pi/4$ (i.e., $z \to -\infty$)
overlapping with those three regions, and then in the remaining regions
by using the connection rules. The result is
\begin{eqnarray}
\label{plus}
\Psi_+^E(Z) &\sim& {C(\hbar)\over P_E(Z)^{1/2}}\; 
\e^{+i\phi(0,Z)/\hbar}\qquad\qquad
\mbox{in the regions ${\cal S}'$, ${\cal S}_1$, ${\cal S}_2$}\nonumber\\
\Psi_+^E(Z) &\sim& {C(\hbar)\over P_E(Z)^{1/2}}\; \e^{-\pi E/\, 2\hbar}
\left( \e^{+i \phi(Z_+,Z)/\hbar} +i\,\e^{-i \phi(Z_+,Z)/\hbar}
\right)\qquad\mbox{in ${\cal S}_0$} \\
\Psi_+^E(Z) &\sim& {C(\hbar)\over P_E(Z)^{1/2}}\; \e^{+\pi E/\,2\hbar} 
\left( \e^{+i \phi(Z_-,Z)/\hbar} -i\,\e^{-i \phi(Z_-,Z)/\hbar}\right)
\qquad\mbox{in ${\cal S}_3$}. \nonumber
\end{eqnarray}
The overall normalization factor $C(\hbar)$ is determined by comparison
with the direct saddle-point evaluation of the integral (\ref{int}):
\begin{equation}
C(\hbar) = K(2\pi\hbar)^{1/4}\e^{-{\pi E/\,4\hbar}} \e^{-i\pi /8} 
\left({E/ \e}\right)^{iE/\,2\hbar}.
\end{equation}

Eqs. (\ref{plus}) readily yield the WKB expansions for the general solution as
\begin{eqnarray}
\label{pm}
\Psi_\lambda^E(Z) &\sim& {C(\hbar)\over P_E(Z)^{1/2}}
\left[\e^{+i \phi(0,Z)/\hbar}+\lambda \e^{-i \phi(0,Z)/\hbar}\right]
\qquad\qquad\qquad\quad \mbox{in ${\cal S}'$} \nonumber\\
\Psi_\lambda^E(Z) &\sim& {C(\hbar)\over P_E(Z)^{1/2}}
\left[\e^{+i \phi(0,Z)/\hbar} + (\lambda-c_-)\e^{-i\phi(0,Z)/\hbar}\right]
\qquad\qquad\mbox{in ${\cal S}_0$} \nonumber\\
\Psi_\lambda^E(Z) &\sim& {C(\hbar)\over P_E(Z)^{1/2}}
\left[(1-\lambda c_+ )\e^{+i \phi(0,Z)/\hbar}
+\lambda \e^{-i \phi(0,Z)/\hbar}\right]
\qquad\quad \mbox{in ${\cal S}_1$} \\
\Psi_\lambda^E(Z) &\sim& {C(\hbar)\over P_E(Z)^{1/2}}
\left[(1-\lambda c_-)\e^{+i \phi(0,Z)/\hbar}
+\lambda\e^{-i \phi(0,Z)/\hbar}\right] \qquad\quad \mbox{in ${\cal S}_2$} \nonumber\\
\Psi_\lambda^E(Z) &\sim& {C(\hbar)\over P_E(Z)^{1/2}}
\left[\e^{+i \phi(0,Z)/\hbar} + (\lambda-c_+)\e^{-i\phi(0,Z)/\hbar}\right]
\qquad\qquad \mbox{in ${\cal S}_3$}, \nonumber
\end{eqnarray}
with the notations
\begin{equation}
\label{cpm}
c_\pm = -\e^{\pm i \pi \nu} = \pm i\,\e^{\pm\pi E/\hbar}
\quad (c_-=1/c_+).
\end{equation}

\section{Large values of the Husimi density}

\subsection{In the WKB framework}
We study the particular solution $\Psi_+^E(Z)$ for a fixed positive energy
$E$ as an example. From the WKB
expansions (\ref{plus}), we derive the Husimi density of this solution, using
the hyperbolic angle as variable. 
In the regions ${\cal S}'$, ${\cal S}_1$, ${\cal S}_2$, away from the turning points, we obtain
\begin{equation}
\label{husimi}
{\cal H}_+^E(Z,\overline Z) \approx 
{|C(\hbar)|^2\over [E(\cosh 2\Re(\theta)+\cos 2\Im(\theta))]^{1/2}}\:
\exp\left\{{E\over\hbar}\Bigl(
-\cosh 2\Re(\theta)[\sin 2\Im(\theta)+1]+\cos 2\Im(\theta)-2\Im(\theta)
\Bigr)\right\}.
\end{equation}
This formula shows that the Husimi measure concentrates semi-classically 
along the maxima of the exponential factor. 
Since the variable $\theta$ is restricted to the strip $|\Im(\theta)|<\pi/2$, 
those maxima occur on the line $\Im(\theta)=-\pi /4$, which corresponds exactly to the branch of hyperbola of energy $E$ in the half-plane $\Im(Z)<0$. 
In the region ${\cal S}_0 \cap \{\Im (Z)\le 0\}$, ${\cal H}_+^E(Z,\overline Z)$ also obeys 
eq. (\ref{husimi}) up to exponentially small terms, 
so that the discussion concerns the whole $E$-hyperbola branch 
in the lower $Z$-half-plane.
The above expression simplifies around this maximum curve, according to the following remarks. First of all, the variables $(Z,\overline Z)$ are 
(up to a factor $-i$) symplectic transforms of the original variables $(q,p)$,
so the expression of the classical energy $E=-{1\over 2}{\overline Z}^2 \,+\,V(Z)$,
where $V(Z)$ is analytic, implies the following classical velocity along the
$E$-energy curve :
\begin{equation}
\dot Z=i{\partial E\over\partial \overline Z}=-i\overline Z
\end{equation} 
and along this curve, we also have $|Z|^2=E\cosh\,2\Re(\theta)$. 
Furthermore, if we decompose a small variation $\delta\theta$ as
\begin{eqnarray}
\delta Z_\parallel&=&\sqrt{2E}\cosh\theta\ \Re(\delta\theta)\nonumber\\
\delta Z_\perp&=&i\sqrt{2E}\cosh\theta\ \Im(\delta\theta)
\end{eqnarray}
($\delta Z_\perp$ is a variation of $Z$ perpendicularly to the $E$-hyperbola), 
then we obtain the following expression of the Husimi density around this maximum curve:
\begin{equation}
\label{husimi1}
{\cal H}_+^E(Z,\overline {Z})\,\approx\, |K|^2\sqrt{2\pi\hbar}\,{1\over|\dot Z|}
\e^{-2|\delta Z_\perp|^2 /\hbar};
\end{equation}
that is, the density decreases as a Gaussian of constant width normally to the maximum curve, its height being given by the inverse of the phase-space
velocity. This corresponds semi-classically to a conserved probability flux
along this curve, and confirms earlier predictions \cite{taka} 
(cf. fig. 1, top left and bottom right).

As a new feature, by contrast, in the upper $Z$-half-plane 
there is a maximum curve in the region ${\cal S}_0$ only, 
and well below the anti-Stokes line where the zeros of $\Psi_+^E(Z)$ lie. 
This maximum curve is given by $\Im(\theta)=+\pi /4$, 
i.e., it is the other branch of the $E$-hyperbola. 
Around it, the Husimi density behaviour is precisely eq. (\ref{husimi1})
times the constant factor $\exp(-{2\pi E/\hbar})$, 
an exponentially small contribution compared to that from the lower half-plane;
hence this enhancement is semi-classically ``invisible" but can be guessed on
the log-plot in fig. 1, top right.
(The correspondence between the $z$ and $Z$ variables is recalled on 
Fig. 2, right.)

\subsection{Energies close to zero}
Now using the expansions (\ref{dexp}), we can analyze the large values of the Husimi density in the case where $E/\hbar$ stays bounded, still in the semi-classical limit $\hbar\to 0$. If we still consider the function $\psi_+^E(z)$, we find a concentration along an invariant subset 
of the separatrix, i.e., the positive real and the imaginary $z$-axes. 
More precisely, if $|z|^2\gg\hbar$, we have
\begin{eqnarray}
\label{expsep}
{\cal H}_+^E(z,\overline z) &\sim& |K|^2\sqrt{2\pi\hbar}\;
{1\over |\dot z|} \e^{-2E\arg(z)/\hbar} \e^{-2\Im(z)^2/\hbar}
\quad\qquad\mbox{when } |\arg(z)|<{\pi\over 4} \\
{\cal H}_+^E(z,\overline z) &\sim& 
{|K|^2\over\cosh(\pi E/\hbar)} \sqrt{\pi\hbar\over 2}\;
{1\over|\dot z|} \e^{2E\arg(-z)/\hbar} \e^{-2\Re(z)^2/\hbar}
\mbox{ when } |\arg(-z)|<{3\pi\over 4}.\nonumber
\end{eqnarray}
We notice that both the longitudinal dependence of the density,
and its Gaussian decrease away from the separatrix, are exactly the same
as for a regular energy curve. Thus, away from the unstable fixed point $z=0$,
the singular limit $E\to 0$ behaves straightforwardly. 
Here, however, the Husimi density is also described exactly
all the way down to the saddle-point (for $z\approx 0$ when $E\approx 0$),
by the explicit formulae (\ref{husi}), e.g.,
\begin{equation}
{\cal H}_+^E(0,0)\,=\,{|K|^2\over\sqrt{8\pi}}\;
\Bigl| \Gamma \Bigl( {1\over 4}+i{E\over 2\hbar}\Bigr) \Bigr| ^2 .
\end{equation}

We can likewise obtain the rough shape of the Husimi density 
for a general solution $\psi_\lambda^E$. As before,
whatever type of expansion we use, we find along each of the four half-axes (which are asymptotes to the classical energy curves)
\begin{equation}
{\cal H}_\lambda ^E (z,\overline {z}) \approx 
I\, {1\over |\dot z|} \e^{-2|\delta Z_\perp|^2 /\hbar}
\end{equation}
where the constant $I$, depending on $\lambda$ and on the half-axis we consider,
can be interpreted semi-classically as the invariant intensity 
of a flux of particles moving with velocity $\dot z=\overline z$ 
along this branch of classical curve. 
For $E \gg 0$, the flux is separately conserved 
along each of the two hyperbola branches: with obvious notations,
\begin{eqnarray}
I_+ &=& I_{-i}\ =\ |K|^2\sqrt{2\pi\hbar}\\
I_- &=& I_{+i}\ =\ |K|^2|\lambda|^2\sqrt{2\pi\hbar} . \nonumber
\end{eqnarray}
When the energy approaches its critical value 0, the intensities become
\begin{eqnarray}
I_+\;&=&\;|K|^2\;\sqrt{2\pi\hbar}\\
I_-\;&=&\;|K|^2\;\sqrt{2\pi\hbar}\;|\lambda|^2\nonumber\\
I_{+i}\;&=&\;{|K|^2\;\sqrt{2\pi\hbar}\over 2\cosh (\pi E/\hbar )} \left\{|\lambda|^2 \e^{\pi E/\hbar} + 2\Im (\lambda)
+ \e^{-\pi E /\hbar}\right\}\nonumber\\
I_{-i}\;&=&\;{|K|^2\;\sqrt{2\pi\hbar}\over 2\cosh (\pi E/\hbar )} \left\{|\lambda|^2 \e^{-\pi E/\hbar} - 2\Im (\lambda)
 + \e^{\pi E/\hbar}\right\}. \nonumber
\end{eqnarray}
Now the flux is only conserved globally:
it is easy to check that $I_{+i} + I_{-i}=I_+ + I_-$.

\section{Asymptotic study of the zeros}
The asymptotic geometry of the zeros for a general solution can be deduced from
the following general principles.
When a function $f(Z)$ has an asymptotic expansion (within a sector) combining 
two exponential behaviours, the function can only vanish 
when both exponential factors are of the same order of magnitude.
Thus, the two exponents must have equal real parts: this necessary condition
defines the anti-Stokes lines of the problem in the complex $Z$-plane.
Zeros of the function can then only develop along anti-Stokes lines 
(always in the large-$|Z|$ approximation), 
and provided both exponentials are present in the given sectorial expansion,
with fixed non-zero prefactors.

\subsection{Energies close to zero - general case}
Here the expansions (\ref{dexp}) have to be used,
and it is simpler to work in the rotated $Z$ variable.
Then the exponential factors read as
 $\e^{+iZ^2/ \,2\hbar}$ and $\e^{-iZ^2/\, 2\hbar}$,
and the anti-Stokes lines on which they balance each other, 
namely the bisecting lines $L_j$ of the quadrants $S_j$,
 are simply the real and imaginary $Z$-axes (fig. 2, right).
Now, for each fixed $j$, two independent solutions 
$\Phi^{(j)}_\pm$ of eq. (\ref{barrier}) can be specified 
by imposing single-exponential asymptotic behaviours along $L_j$, as
$\Phi^{(j)}_\pm \sim f^{(j)}_\pm(Z) \e^{\pm iZ^2/\, 2\hbar}$.
A general solution $\Psi(Z)$ is then proportional to 
$\left[ \Lambda \Phi^{(j)}_+ + \Phi^{(j)}_- \right]$ 
for some $\Lambda \in \overline \Bbb C$, and satisfies
\begin{equation}
\Psi(Z) \propto 
\Lambda f^{(j)}_+(Z) \e^{+iZ^2/\, 2\hbar}+ f^{(j)}_-(Z) \e^{-iZ^2/\, 2\hbar},
\qquad Z\to \infty \hbox{ in } S_j ,
\end{equation}
whence the condition $\Psi(Z)=0$ in the sector $S_j$ asymptotically reads as
\begin{equation}
\label{zeq}
Z^2/\hbar \sim (-1)^j 2\pi m + i\log\left[\Lambda \, r^{(j)}(Z)\right], \quad
\mbox{ for } m\to +\infty, \quad r^{(j)}(Z) \equiv -f^{(j)}_+(Z)/f^{(j)}_-(Z).
\end{equation}
This yields an asymptotic sequence of zeros, as:
$Z^{(j)}_m \sim \e^{i j\pi/2}\sqrt{2\pi m \hbar}$ to leading order; 
thereupon, to the order $O(1)$ included,
\begin{equation}
\label{zer}
Z^{(j)\,2}_m/\hbar \sim 
(-1)^j 2 \pi m + i \log\ r^{(j)}(\e^{i j\pi/2}\sqrt{2\pi m \hbar}) 
+ i \log\Lambda +O(m^{-1}\log m), \qquad m\to +\infty .
\end{equation}
The final square-root extraction is straightforward,
\begin{equation}
Z^{(j)}_m/\sqrt\hbar \sim \e^{i j\pi/2}\left(\sqrt{2\pi m}
+(-1)^j i \:{\log\ r^{(j)}(\e^{i j\pi/2}\sqrt{2\pi m \hbar}) 
+\log \Lambda \over 2 \sqrt{2 \pi m}} 
+O\left({\log^2 m\over m^{3/2}}\right) \right)
\end{equation}
so that the simpler form (\ref{zer}) will be preferred for further displays of results.

When the asymptotic analysis concerns a fixed linear combination 
given as $[\lambda D_{\nu}(y)+D_{\nu}(-y)]$, $\Lambda$ turns into 
a sector-dependent function $\Lambda^{(j)}(\lambda)$, which is 
the linear fractional transformation induced by the change of basis
$\{D_\nu(y),D_\nu(-y)\} \longrightarrow \{\Phi^{(j)}_+(Z),\Phi^{(j)}_-(Z)\}$.
The equations for zeros like (\ref{zer}) become singular for the two values 
of $\lambda$ which map to $\Lambda^{(j)}=0 \hbox{ or } \infty$, 
simply because they yield the pure $\Phi^{(j)}_-$ or $\Phi^{(j)}_+$ solutions
which have no zeros (at least asymptotically) in this sector $S_j$.

\medskip
We now list more explicit results. In the sector $S_0$,
corresponding to $\{0<\arg y<+\pi/2\}$,
\begin{equation}
\lambda D_{\nu}(y)+D_{\nu}(\e^{-i\pi} y) \sim 
(\lambda + \e^{-i\pi\nu}) y^\nu \e^{-y^2/4}
+{\sqrt{2\pi} \over \Gamma(-\nu)} y^{-\nu-1}\e^{+y^2/4} .
\end{equation}
Upon the substitutions $\Lambda^{(0)}(\lambda)r^{(0)}(Z) \equiv
\bigl(\sqrt{2\pi}/\Gamma(-\nu)\bigr) y^{-2\nu-1}/(\lambda + \e^{-i\pi\nu}),
\ y=\sqrt{2/\hbar}\,\e^{i\pi/4}Z$ and $\nu=-{1 \over 2}-{iE \over \hbar}$, 
the asymptotic equation (\ref{zer}) for zeros becomes
\begin{equation}
\label{zr2}
{Z^{(0)\,2}_m \over \hbar} \sim (2m-1)\pi -{E \over\hbar} \log\, 4m\pi i
 -i \,\log { \Gamma({1 \over 2} +i{E \over \hbar}) \over \sqrt{2\pi}}
-i \,\log (\lambda + i \e^{-\pi E/\hbar}) , 
\qquad -{\pi \over 4}<\arg Z<+{\pi \over 4}.
\end{equation}

A geometrical interpretation will prove useful. Let $C_\pm$ be the two circles 
in the $\lambda$-plane respectively specified by the parameters 
(cf. eq. (\ref{cpm}) and fig. 3)
\begin{equation}
\label{circ}
\mbox{centers:}\quad c_\pm = \pm i\,\e^{\pm\pi E/\hbar},
\qquad \mbox{radii:} \quad R_\pm=(1+\e^{\pm 2\pi E/\hbar})^{1/2} 
\end{equation}
(they are both centered on the imaginary axis
and intersect orthogonally at $\lambda=+1$ and $-1$).
Let us also write, for real $t$,  $\Gamma({1\over 2} + it)/\sqrt{2\pi}$ 
in polar form as $(2 \cosh \pi t)^{-1/2} \e^{i \Theta(t)}$,
defining the phase $ \Theta(t)$ by continuity from $ \Theta(0)=0$.
The formula (\ref{zr2}), and its partners in the other sectors, then become:

\bigskip
\noindent in $S_0=\{-{\pi /4}<\arg Z<+{\pi /4}\}$,
\begin{equation}
\label{z0}
{Z^{(0)\,2}_m \over \hbar} \sim (2m-1)\pi -{E \over\hbar} \log\ 4m\pi 
+ \Theta( {E/\hbar} ) -i \,\log { \lambda -c_- \over R_-}
\quad (\lambda \notin \{c_-, \infty\})
\end{equation}
in $S_1=\{+{\pi /4}<\arg Z<+{3\pi /4}\}$,
\begin{equation}
\label{z1}
-{Z^{(1)\,2}_m \over \hbar} \sim (2m-1)\pi +{E \over\hbar} \log\ 4m\pi 
- \Theta( {E/\hbar} ) +i \,\log { \lambda^{-1} -c_+ \over R_+}
\quad (\lambda \notin \{0,c_-\}) 
\end{equation}
in $S_2=\{-{5\pi /4}<\arg Z<-{3\pi /4}\}$,
\begin{equation}
\label{z2}
{Z^{(2)\,2}_m \over \hbar} \sim (2m-1)\pi -{E \over\hbar} \log\ 4m\pi 
+ \Theta( {E/\hbar} ) -i \,\log { \lambda^{-1} -c_- \over R_-}
\quad (\lambda \notin \{0,c_+\}) 
\end{equation}
in $S_3=\{-{3\pi /4}<\arg Z<-{\pi /4}\}$,
\begin{equation}
\label{z3}
-{Z^{(3)\,2}_m \over \hbar} \sim (2m-1)\pi +{E \over\hbar} \log\ 4m\pi 
- \Theta( {E/\hbar} ) +i \,\log { \lambda -c_+ \over R_+}
\quad (\lambda \notin \{c_+, \infty\}) .
\end{equation}

For general values of the parameters we cannot get more precise information
this way, except by going to higher orders (but still in the asymptotic sense). 
We have yet no idea about the position, or even existence, of small zeros. 
Our subsequent strategy will be 
to start from very special cases for which the pattern of zeros is well known,
and from there to vary continuously the parameters $E$ and $\lambda$ 
and keep track of the zeros along these deformations: zeros are topological defects, so they move continuously w.r. to both parameters.
We will then exploit symmetries of eq. (\ref{barrier}), 
especially the reality of its solutions (a real function 
of a complex variable $t$ is one satisfying $f(\overline t)=\overline{f(t)}$). 
Eq. (\ref{barrier}) is a differential equation with real coefficients, hence it
admits real solutions, whose zeros are symmetrical w. r. to the real $Z$-axis.
During a deformation of a real solution, zeros could then only leave (or enter) 
the real axis in conjugate pairs,
but at the same time the functions considered here 
(solutions of a second-order equation) cannot develop double zeros;
consequently, each zero is bound to stay permanently real (or nonreal) 
in the course of a real deformation.

\subsection{Even--odd solutions of zero energy}
We use the $E=0$ expressions eq. (\ref{bessel}) in terms of the Bessel functions
$J_{\pm 1/4}(Z^2/\,2\hbar)$ to view the pattern of zeros more precisely
in this particular case. 

We know that, for real $\mu$, $t^{-\mu}J_\mu(t)$ is a real even function
having only real zeros $\pm j_{\mu,m},\ m=1,2,\cdots$, with $ j_{\mu,m}>0$
and $ j_{\mu,m} \sim \pi (m+{\mu\over 2} - {1\over 4})$ for large $m$.
This translates into the $Z$ variable as follows.

For $\lambda =+1$: the even solution (fig. 1, middle right)
\begin{equation}
\label{bessele}
\Psi_{+1}^0(Z)\,=\,K{\left({2\pi^3/\hbar}\right)}^{1/4} Z^{1/2}\; J_{-1/4}\left({Z^2\over 2\hbar}\right)
\end{equation} 
(a real function for $K$ real), 
is not just even but also invariant under $Z \longrightarrow iZ$;
all its zeros are purely real or imaginary,
the $m$-th positive zero admits the approximation
\begin{equation}
\label{zeven}
Z_m^{(0)}|_{\lambda =+1} \sim \sqrt{\hbar(2m\pi - 3\pi/ 4)},
\end{equation}
and all other zeros follow by the rotational symmetry of order 4.

For $\lambda =-1$: the odd solution (fig. 1, bottom left)
\begin{equation}
\label{besselo}
\Psi_{-1}^0(Z)\,=\,K\,e^{i\pi/ 4}\,{\left({2\pi^3/\hbar}\right)}^{1/4} Z^{1/2}\;J_{1/4}\left({Z^2\over 2\hbar}\right)
\end{equation} 
has exactly the same symmetries for its zeros as the even one
(because the auxiliary function $\e^{-i\pi/ 4}\Psi_{-1}^0(Z)/Z$ 
has all the symmetries of $\Psi_{+1}^0(Z)$);
besides an obvious zero at the origin,
the $m$-th positive zero of $\Psi_{-1}^0(Z)$ lies approximately at
\begin{equation}
Z_m^{(0)}|_{\lambda =-1} \sim \sqrt{\hbar(2m\pi - \pi/ 4)}.
\end{equation}

\subsection{Even--odd solutions of non-zero energy}
 If we ``switch on" the energy, keeping $\lambda=+1$, the solutions will not 
be Bessel functions any longer, but they still exhibit interesting features.
Equation (\ref{barrier}) has real coefficients, 
hence its even and odd solutions are real up to a constant factor, 
i.e., up to an adjustment of $\arg K$.
For the even solution $\Psi_{+1}^E(Z)$, the reality condition is
\begin{equation}
\label{Keven}
\Psi_{+1}^E(0)\,=\,K{\left(2\over\pi\right)}^{1/4}(2\hbar)^{iE\over 2\hbar}\;
\Gamma\left({1\over4} + {iE\over 2\hbar}\right) \quad\mbox{real.}
\end{equation}  
This even solution can be seen as a real deformation 
of the solution (\ref{bessele}). 
It maintains the symmetry w.r. to the origin and to the two $Z$ coordinate axes;
only the $\pi/2$ rotation symmetry is lost for $E\ne 0$.
Due to the two mirror symmetries, as explained above,
the only possible motion of the zeros during this deformation is
a ``creeping without crossing" along the four half-axes, 
symmetrically w.r. to the origin. 
This can be checked on the $O(1)$ terms of the expansions (\ref{z0}--\ref{z3}),
and on the sequence of plots: fig. 1 (middle right), fig. 6 (right), 
fig. 1 (bottom right). 
At the same time, this deformation allows to properly count the zeros 
at all energies, by continuity from $E=0$
where eq. (\ref{zeven}) does count the zeros:
in each of the expansions (\ref{z0}--\ref{z3}),
$Z^{(j)}_m$ remains the actual $m$-th zero on the half-axis $L_j$
if the corresponding complex log functions are defined at $\lambda=+1$ as
\begin{equation}
\log (1-c_+) = \log |1-c_+| - i \,\arctan \e^{+\pi E/\hbar}, \qquad
\log (1-c_-) = \log |1-c_-| + i \,\arctan \e^{-\pi E/\hbar}
\end{equation}
where the $\arctan$ function has the usual range $(-{\pi/ 2},{\pi/ 2})$.

The same analysis can be performed for the odd real solution $\Psi_{-1}^E(Z)$
 of (\ref{barrier}), a deformation of the solution (\ref{besselo}),
for which the reality condition is
\begin{equation}
\label{Kodd}
{d\Psi_{-1}^E\over dZ}(0)\,=\,K{\left({2^7\over\pi}\right)}^{1/4}(2\hbar)^{-{1\over 2}+{iE\over 2\hbar}}\;\;
\Gamma\left({3\over4} + {iE\over 2\hbar}\right)\, \e^{i\pi/ 4}
\quad\mbox{real.}
\end{equation}

\subsection{Real solutions}
We now consider more general families of real solutions, i.e.,
$\Psi_\lambda ^E(\overline {Z}) =\overline{\Psi_\lambda ^E(Z)},\,\forall Z\in \Bbb C$. These exist only for certain values of $\lambda$, for which we 
have to adjust $K$. Since (\ref{barrier}) is a second-order equation with
real coefficients, 
it has the real solutions 
\begin{equation}
\Psi^E(Z) = \kappa\left( \Psi_{+1}^E(Z)+t \Psi_{-1}^E(Z)\right), \quad\mbox{for}\quad
\kappa \in \Bbb R^\ast,\quad t\in \Bbb R\cup\{\infty\} = \overline \Bbb R .
\end{equation}
Under the change of basis 
$\{\Psi_{+1}^E(Z),\Psi_{-1}^E(Z)\}\longrightarrow \{D_\nu(y),D_\nu(-y)\}$, 
the set $\{t\in\overline \Bbb R\}$ is mapped to a circle in the projective 
$\lambda$-plane, passing through $\lambda|_{t=0}=+1,\ \lambda|_{t=\infty}=-1$.
The full circle can be determined long-hand 
using eqs. (\ref{Keven}--\ref{Kodd}),
but more easily by asking the expansions (\ref{z0}), (\ref{z2}) to yield
asymptotically real zeros as reality demands:
the resulting $\lambda$-circle is $C_-$ 
(note that $\lambda \in C_- \Longleftrightarrow 1/\lambda \in C_-$).

Let us now analyze the motion of the zeros in the four sectors 
as we vary $\lambda$ around $C_-$ anti-clockwise from 1 to $\e^{2i\pi}$. 
We already know that the real zeros cannot leave the real axis by symmetry. 
In the expansions (\ref{z0}) and (\ref{z2}) for $S_0$ and $S_2$ respectively, 
the only modifications are that $\arg(\lambda - c_-)$ increases by $2\pi$, 
and $\arg(\lambda^{-1} - c_-)$ decreases by $2\pi$, inducing 
a re-labeling of the large zeros in those two sectors.
Hence each large positive or negative zero creeps to the right 
until it reaches the former position of its right neighbour after one cycle.
The small zeros, trapped on a real bounded interval in-between by reality,
and unable to cross one another,
can then only follow the same homotopic pattern of behaviour. 

 In each of the two remaining sectors, by contrast,
the zeros are not confined to the imaginary axis
whereas the cycle $C_-$ is homotopically trivial 
in the Riemann surface of the relevant logarithmic function.
According to the expansions in $S_1$  (resp. $S_3$), 
the large zeros in these sectors
perform a clockwise (resp. anti-clockwise) cycle beginning and ending 
at their location on the imaginary axis for $\lambda=1$. 
The geometric relation $\overline{\lambda -c_+}=R_+^2(\lambda ^{-1}-c_+)^{-1}$ 
is the asymptotic remnant of reality: $\{\Psi(Z)=0\Rightarrow \Psi(\overline Z)=0\}$.
This motion in $S_1$ and $S_3$ has been thus shown only for large zeros, 
but we may argue a similar behaviour for the smaller ones by homotopy;
moreover, by reality these zeros cannot cross the real axis and
are confined to a bounded region around the origin (contrary to the real zeros,
they cannot migrate to infinity when $\lambda$ keeps revolving around $C_-$); orbits of non-real zeros during the described $\lambda$-cycle are 
also symmetrical w.r. to the imaginary axis, 
thanks to a $\lambda \leftrightarrow 1/\lambda$ symmetry of the real solutions.

The evolution of all the zeros under the $\lambda$-cycle $C_-$ is globally
depicted for $E=0$ on fig. 4 (left).
  
 Similarly, the solutions corresponding to $\lambda\in C_+$ 
can be chosen real w.r. to the variable $iZ$ (observing that eq. (\ref{barrier})
is invariant under the change $\{Z\to iZ, E\to -E\}$), and analyzed likewise.  

\subsection{Singular values of $\lambda$}
\label{sing}
Each of the expansions (\ref{z0}--\ref{z3}) becomes singular
for two values of $\lambda$ from the set $\{0, \infty, c_-,c_+\}$. 
The solutions are then $\psi^E_\pm(z), \psi^{-E}_\pm(iz)$, 
corresponding to ``pure $D_\nu$ functions",
and eqs. (\ref{dexp}) clearly show that $D_\nu(y)$ has large zeros only in two
out of four sectors, namely along the pair of adjacent anti-Stokes lines
$\{\arg y = \pm 3\pi/4\}$. That is why
the zeros in the two other sectors must ``escape to infinity" 
as $\lambda$ moves towards one of its special values. 
Let for instance $\lambda$ decrease from $+1$ (the even solution case) 
to the special value $0$ along the interval $[0,1]$. 
For simplicity we first restrict ourselves to the case $E=0$, 
where the eigenfunctions are combinations of Bessel functions. 
The zeros' expansions (\ref{z0}--\ref{z3}) 
(with $R_\pm |_{E=0}=\sqrt 2$) then become singular in the sectors
$S_1$ and $S_2$, whereas they stay perfectly uniform in the two other sectors.
In $S_2$, eq. (\ref{z2}) holds uniformly as $Z^{(2)\,2}_m /\hbar \to \infty$,
which amounts to $-\log \lambda + 2 \pi m \gg 1$, 
and the same conclusion is reached for eq. (\ref{z1}) in $S_1$. 
As $\lambda \to +0$, the two formulae merge into a single one,
\begin{equation}
Z_n^2/\hbar \sim i \,\log {\lambda \over \sqrt 2} +(2n-1) \pi 
\qquad \mbox{in } S_1\cup S_2 \quad\mbox{for any }n \in \Bbb Z
\end{equation}
where the global counting index $n \in \Bbb Z$ matches with $m$ in the sector
$S_2$ for $n \gg 0$ and with $(-m+1)$ in $S_1$ for $n \ll 0$. 
These zeros thus tend to follow the hyperbola branch 
$\{2\Re(Z) \Im(Z) = \hbar \log \ \lambda /\sqrt 2,\ \Re(Z)<0\}$,
which itself recedes to infinity as $\lambda \to +0$, as can be seen on
fig. 4 (right) followed by fig. 1 (middle left).

This description can be generalized to the case of a non-vanishing
(but small) energy. The asymptotic condition for zeros in $S_1 \cup S_2$
when $\lambda\to 0$ must now be drawn from eq. (\ref{zeq}) itself and reads as
\begin{equation}
{Z_n^2\over\hbar} \approx (2n-1)\pi 
-i\,\log\left({\lambda^{-1}-c_\mp\over R_-}\right)
-{E\over\hbar} \log\left(2 {Z_n^2\over\hbar}\right) + \Theta( {E/\hbar} )
\qquad \mbox{for } n \to \pm \infty .
\end{equation}
For small values of $\lambda$, these zeros tend to follow asymptotically 
two half-branches of hyperbolae which differ as $n \to +\infty$ or $-\infty$
(because of the $\Im (\log Z_n^2)$ contribution);
the matching between this set of zeros for small $\lambda$ and the set of zeros
of $\Psi_{+1}^E$ on $\Bbb R_-\cup i \Bbb R_+$ can be done as in the case $E=0$. 

\subsection{WKB expansions of zeros at a fixed non-vanishing energy }
We first consider the particular solution $\Psi_+^E(Z)$.
In the semi-classical limit of eqs. (\ref{plus}), $\Psi_+^E(Z)$ can vanish only
within the regions ${\cal S}_0$ and ${\cal S}_3$ and along anti-Stokes lines,
now defined by the conditions:
$\phi(Z_\pm,Z)/\hbar \mbox{ real}$, respectively. 
From the complete set of anti-Stokes lines 
(shown on fig. 5 top left, with $\arg \hbar=0$ henceforth), 
the presently relevant ones are:
in ${\cal S}_3$, the imaginary half-axis below $Z_-$; and in ${\cal S}_0$, 
the anti-Stokes line from $Z_+$ asymptotic to the positive real axis
(fig. 5, bottom right).
The zeros themselves are given asymptotically by the equations
 
\begin{eqnarray}
\label{phi}
\phi(Z_+,Z)&=&(+m-1/4)\hbar\pi,\quad m\in\Bbb N,\ m \gg 1 \qquad\mbox{in ${\cal S}_0$}
\nonumber\\
\phi(Z_-,Z)&=&(-m+1/4)\hbar\pi,\quad m\in\Bbb N,\ m \gg 1 \qquad\mbox{in ${\cal S}_3$} .
\end{eqnarray}
Finally, the expansions 
\begin{equation}
\phi(Z_+,Z) \sim +{Z^2 / 2}\quad\mbox{when}\quad Z\to +\infty , \qquad
\phi(Z_-,Z) \sim +{Z^2 / 2}\quad\mbox{when}\quad Z\to -i\infty
\end{equation}
restore the former large zero behaviours, 
$Z_m^{(j)}\sim \e^{ij\pi/2}\sqrt{2\hbar m\pi}$ for $j=0,3$ respectively.

We can also see how the zeros in regions ${\cal S}_1$ and ${\cal S}_2$ go to infinity
for a more general eigenfunction $\Psi_\lambda^E(Z)$, 
when the parameter $\lambda$ decreases from $1$ to $0$ along $[0,1]$ 
as in subsection (\ref{sing}), but now for a fixed non-vanishing energy. 
We will then be able to compare the results in the two frameworks.

Using the WKB expansions (\ref{pm}) in the different regions of the $Z$-plane, 
we study the equation $\Psi_\lambda^E(Z)=0$ in each of them \cite{olv1}. 
For $\lambda=0$ we found zeros along only two anti-Stokes lines. 
For a general value, there will be zeros in all regions
(and, inasmuch as $\lambda$ varies independently of $\hbar$, 
those zeros are not confined near the above anti-Stokes lines). 
Moreover, we know from symmetry properties that
for $\lambda=\pm 1$, the zeros can only lie on the real and imaginary axes. 
It is also obvious, from the different expansions, that the zeros' pattern
depends on the ratio $\lambda /c_- = \lambda c_+$ (cf. eq. (\ref{cpm})).

In the range $1>\lambda\gg |c_-|$, the differences between the formulae
(\ref{pm}) for regions ${\cal S}'$, ${\cal S}_0$, ${\cal S}_2$ are irrelevant 
as far as the position of the zeros is concerned, 
so that the first one suffices to localize the zeros in those three regions
by means of the equations
\begin{eqnarray}
\label{tilde}
\Im \,\phi(0,Z) &=& -(i\hbar/2) \log\lambda \\
\Re \,\phi(0,Z) &=& \hbar {\pi/ 2} \quad {\rm mod}\ \hbar\pi.\nonumber
\end{eqnarray}
For $\lambda =1$, the zeros are exactly real by symmetry.
When $\lambda$ decreases towards $|c_-|$, 
the curve (\ref{tilde}) gets deformed towards $Z_+$, 
keeping the real axis as asymptote at both ends. 
At the same time, the zeros in ${\cal S}_1 \cup {\cal S}_3$ 
stay along the imaginary axis. 
When $\lambda\approx |c_-|$, there are zeros along all anti-Stokes lines
from $Z_+$ and along the imaginary axis in ${\cal S}_3$. 
When $\lambda \ll |c_-|$, the zeros along the anti-Stokes lines in 
${\cal S}_0 \cup {\cal S}_3$ stabilize, 
whereas the ones in the regions ${\cal S}_1 \cup {\cal S}_2$ 
lie along the curve (\ref{tilde}), which recedes to infinity when $\lambda\to 0$
as in the previous subsection, keeping $i\Bbb R_+$ and $\Bbb R_-$ as asymptotes.
All these phenomena appear on the sequence of plots in fig. 5.
Once more, we can recover the previous asymptotic large zeros
by expanding the action integrals $\phi(0,Z)$ along the four half-axes.

\section{Conclusion}
The above study should help to better understand more complicated 1-d systems when the eigenenergy is very near a saddle-point value and standard WKB theory fails.

This can be illustrated upon the eigenstates of a quantum Harper model.
The classical Harper Hamiltonian is $H_{\rm H}=-\cos(2\pi P) - \cos(2\pi Q)$ 
on the torus phase space $(Q \mbox{ mod } 1,P \mbox{ mod } 1)$.
We have quantized it on the Hilbert space of wavefunctions with periodic boundary conditions, 
which has a finite dimension $N$ to be identified with $(2 \pi \hbar)^{-1}$,
and taken $N=31$ for calculations. 
We have then selected the eigenfunction $\Psi_n$ immediately below the separatrix energy $E=0$; it is an even state having the quantum number $n=14$,
and its Husimi density over the torus is plotted on fig. 6, left
(cf. also \cite{leb:vor} 1990, fig. 2a).
Now, by expanding the Hamiltonian near one saddle-point like $(Q=0, P=-1/2)$
we recover a quadratic-barrier problem. This suggests to compare $\Psi_{14}$
with the even dilator eigenfunction $\Psi_{+1}^E(Z)$ 
for $E/\hbar \equiv 2 \pi N |E_{14}|/(2 \pi)^2$
and $z \equiv \hbar^{-1/2}\e^{i\pi/4}(Q-iP)/\sqrt 2 \approx \e^{i\pi/4}10.(Q-iP)$ 
(see fig. 6 right). 
We then observe that not only the high densities of both figures 
around the saddle-points fit very nicely 
(a result to be expected from semiclassical comparison arguments
\cite {taka:hyp,taka}),
but also, more surprisingly, the sequences of zeros for the Harper eigenfunction
match the comparison zeros as well, and not just near the saddle-points but
practically all the way out to the extremal points; at these points the lines
supporting the zeros intersect and the correspondence must fail, 
but this happens much beyond its reasonable range of validity anyway.

\vfill\eject

\centerline{\bf Figure captions}
\bigskip
Fig. 1. Contour plots of the Husimi densities 
${\cal H}^E_\lambda(z,\overline z)$ ($\hbar=K=1$) for selected eigenfunctions
$\psi^E_\lambda$ of the quantum dilator $\hat H$, 
using uniform level spacing $0.2$ (linear plots), except for the top-right plot
(drawn in a logarithmic scale using the uniform level
ratio $1/{\rm e}$ from the maximum downwards). The zeros of the Husimi densities,
constituting the stellar representation, are shown as dots (except in top row).
Maxima in the $E=0$ plots: 
${\cal H}^{E=0}_{+,{\rm max}} \approx 3.6963$ at $z \approx 0.5409$ 
(at middle left);
${\cal H}^{E=0}_{+1}$ (even state, at middle right) very steeply peaks at $z=0$,
with maximum value $ \approx 10.488$;
${\cal H}^{E=0}_{-1}$ (odd state, at bottom left) vanishes at $z=0$ 
and reaches its maximum value $ \approx 2.0054$ 
at the four points $i^j z_0,\ z_0 \approx 1.162$.

\medskip

Fig. 2. Stokes regions for the asymptotic $\hbar \to 0$ expansions of the
eigenfunctions of $\hat H$.
Left: fixed $E$ regime (eqs. (\ref{plus}),(\ref{pm})), with $\arg \hbar = \epsilon >0$ 
(the Stokes lines for $\epsilon \longrightarrow -\epsilon $ 
are the left--right transposed of these);
$Z_\pm$ are the two turning points; cuts for the function $P_E(Z)$ are
conveniently placed over the two thicker Stokes lines.
Right: fixed $E/\hbar$ regime (eqs. (\ref{dexp}),(\ref{z0})--(\ref{z3})); 
this picture is the scaling limit of the preceding one for $E \to 0$
when $\epsilon =0$ (except for the place of cuts).

\medskip

Fig. 3. The complex plane of the coefficient $\lambda$ with the circles
$C_-$ corresponding to $Z$-real solutions and
$C_+$ corresponding to $iZ$-real solutions; the finite special values
$0,c_-,c_+$ are marked (the fourth one being $\lambda = \infty$).

\medskip

Fig. 4. Left: positions of the zeros of $\psi^{E=0}_\lambda(z)$ 
for four values of $\lambda$ in anticlockwise succession on the circle $C_-$:
$\lambda_0=+1$, $\lambda_1=(\sqrt 2 -1)i$, $\lambda_2=-1$, and
$\lambda_3=(-\sqrt 2 -1)i$.
Right: same for the decreasing positive sequence $\lambda_n=10^{-n},\ n=0,3,6$;
fig. 1 (middle left) displays the limiting special value $\lambda=0$.

\medskip

Fig. 5. Top left: the anti-Stokes lines for the semiclassical
expansions of the eigenfunctions of $\hat H$ in the fixed $E$ regime.
Remaining five plots: positions of the zeros of $\psi^{E=+1}_\lambda(z)$,
exceptionally with $\hbar=0.2$, for a decreasing positive sequence of the type:
$\lambda=+1$; $|c_-| \ll \lambda <1$; $\lambda=|c_-|(=\e^{-\pi E/\hbar})$; 
$0<\lambda \ll |c_-|$; $\lambda=0$. The thicker curves are the portions of
the $H=E$ hyperbola where the Husimi measure concentrates semiclassically.

\medskip

Fig. 6. Left: $15^{\rm th}$ eigenstate of the Harper model quantized in
dimension $N=1/(2\pi\hbar) =31$ (with periodic boundary conditions); 
$\Psi_{14}(Z)$ has to have a double zero at $(0,0)$ to yield the correct count
of $14$ zeros within the $\{H_{\rm H}<E\}$ region (and $(31-14)$ outside)
\cite{leb:vor}; 
this Husimi density contour plot exceptionally uses level spacing 0.089 
to match the spacing-to-maximum ratio of the comparison plot (next).
Right: the Husimi density ${\cal H}^E_{+1}$ serving as comparison function,
plotted as in fig. 1 (maximum value $\approx 6.752$, at the origin).

\vfill\eject

\begin{figure}
\epsfxsize 17 truecm
\epsfbox{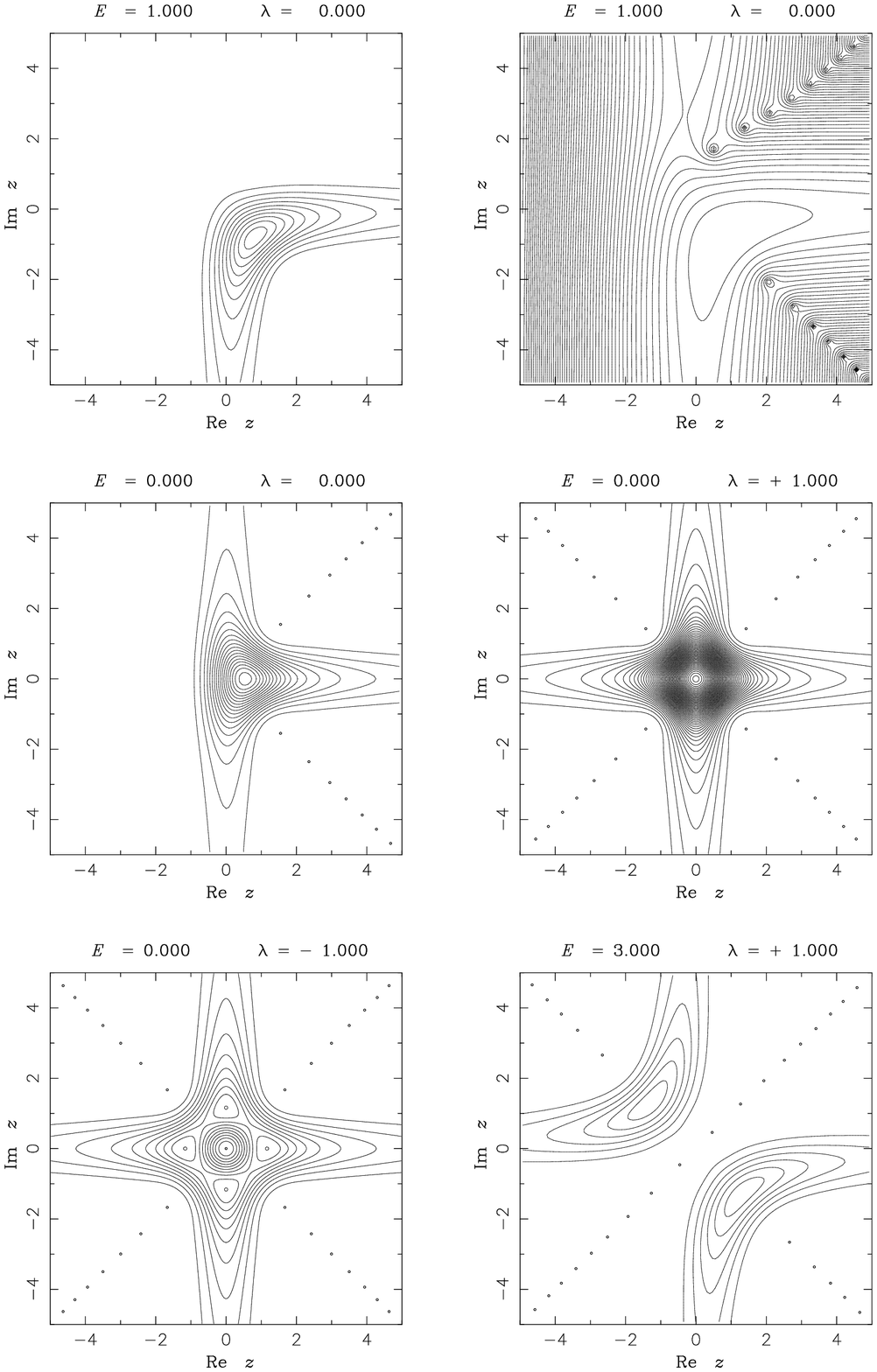}
\caption{}
\end{figure}
\vfill\eject
\begin{figure}
\epsfxsize 17 truecm
\caption{}
\epsfbox{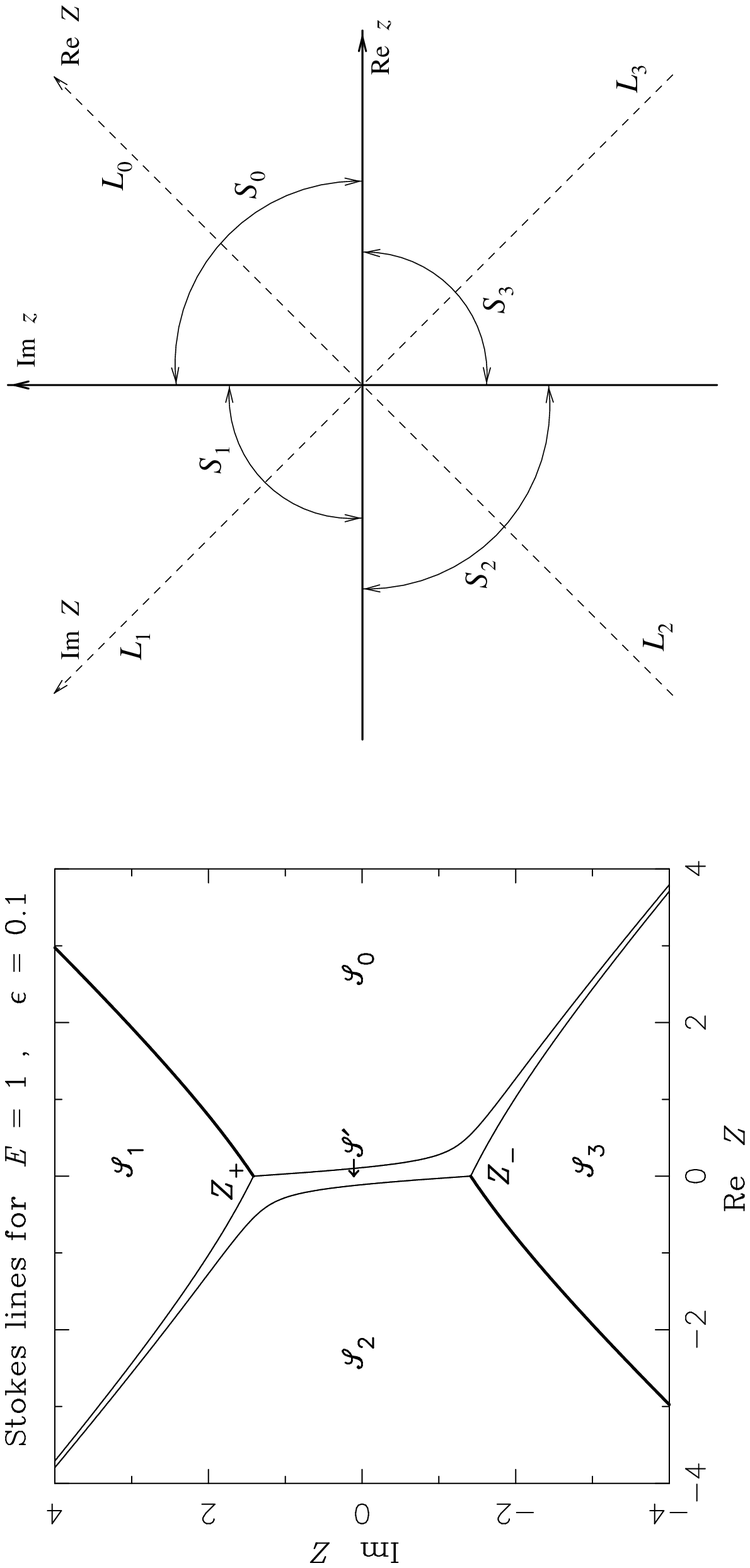}
\end{figure}
\vfill\eject
\begin{figure}
\epsfxsize 17 truecm
\epsfbox{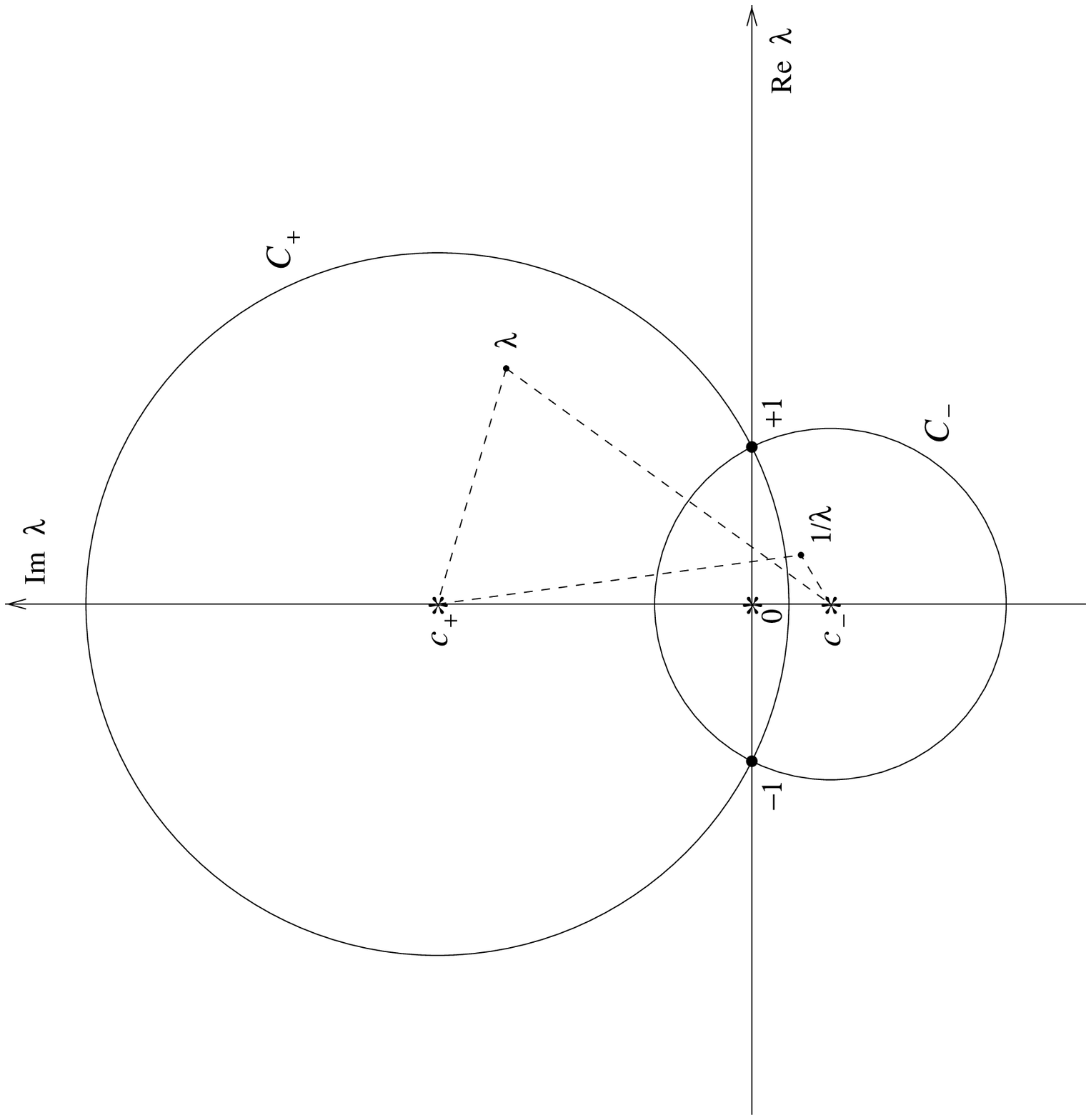}
\caption{}
\end{figure}
\vfill\eject
\begin{figure}
\epsfxsize 17 truecm
\epsfbox{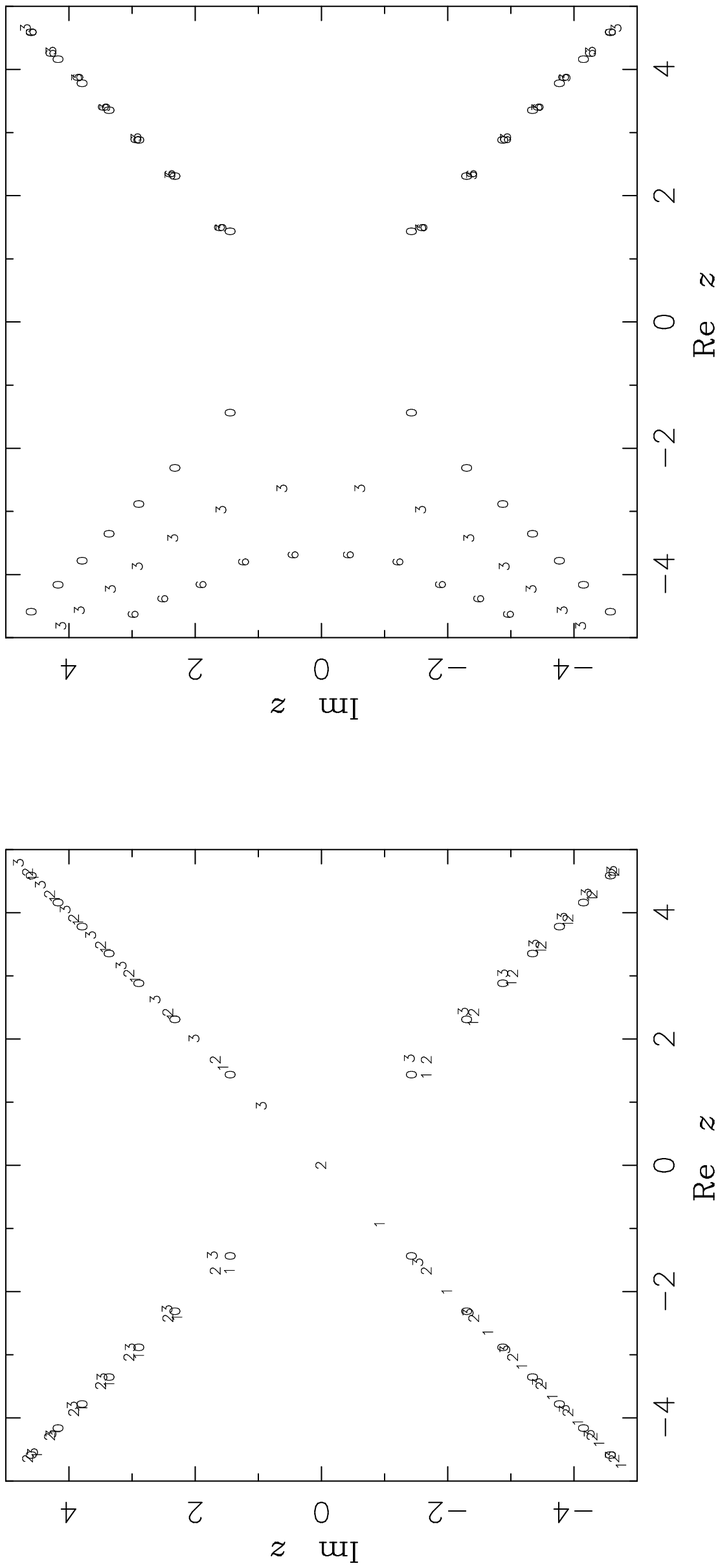}
\caption{}
\end{figure}
\vfill\eject
\begin{figure}
\epsfxsize 17 truecm
\epsfbox{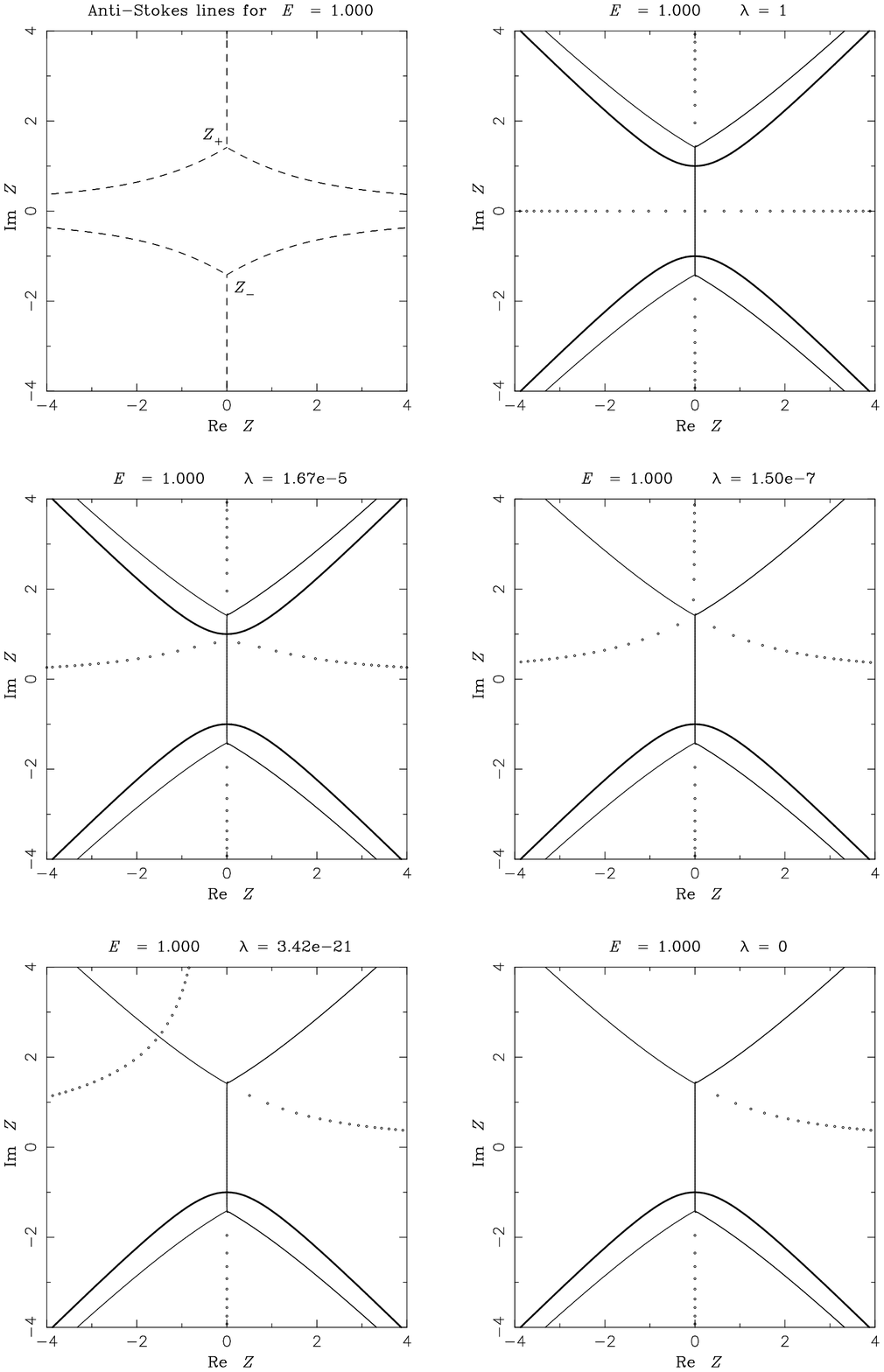}
\caption{}
\end{figure}
\vfill\eject
\begin{figure}
\epsfxsize 17 truecm
\epsfbox{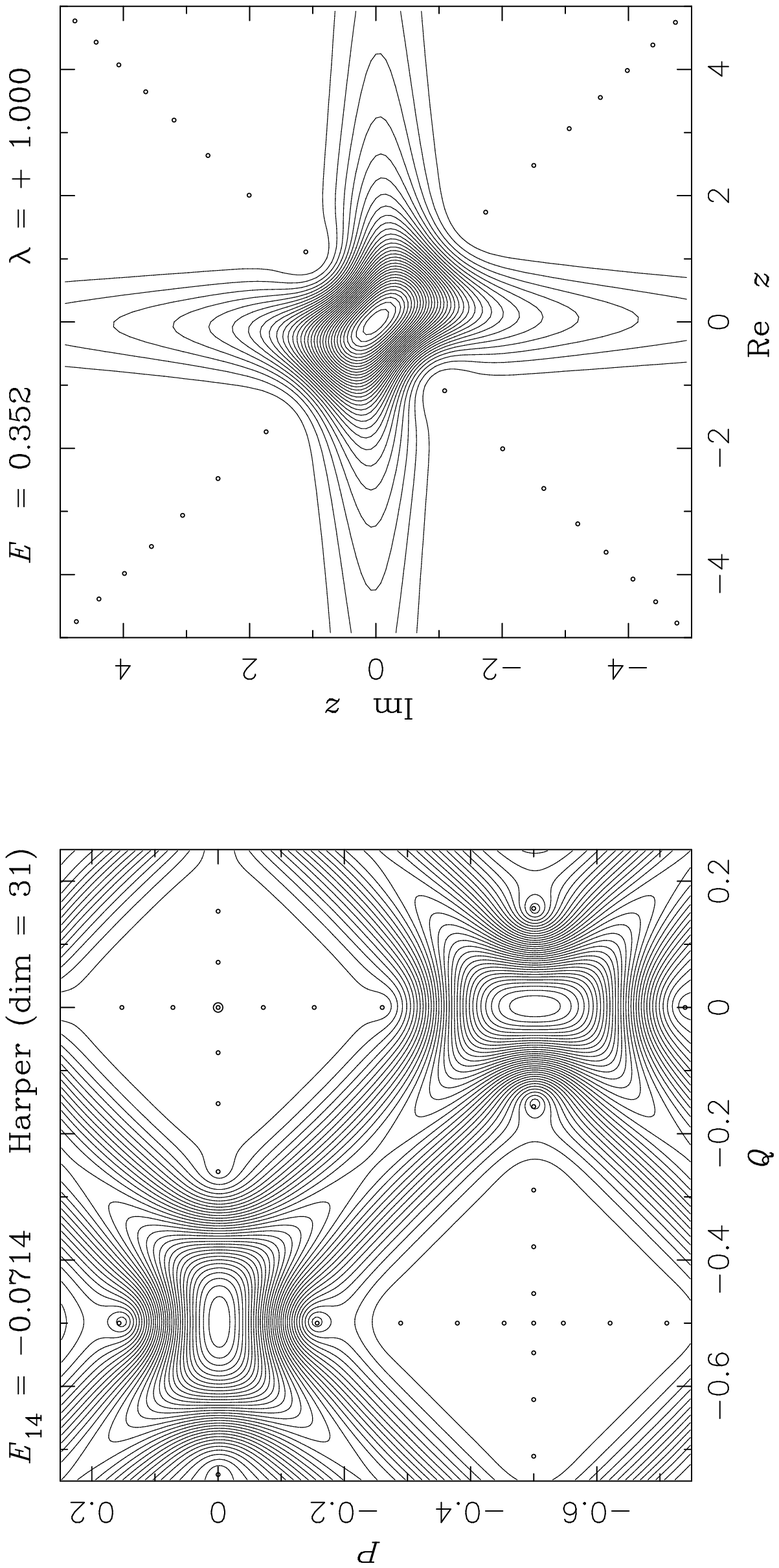}
\caption{}
\end{figure}

\end{document}